\documentstyle[prl,epsf,twocolumn,aps,graphics,graphicx]{revtex}
\newcommand{\sr}{SrB$_6$~}
\newcommand{\ba}{BaB$_6$~}

\begin{document}
\twocolumn[\hsize\textwidth\columnwidth\hsize\csname @twocolumnfalse\endcsname

\title{Electronic Fine Structure in the Electron-Hole Plasma in \sr}

\author{C. O. Rodriguez,$^1$ Ruben Weht,$^2$
          and W. E. Pickett$^3$}

\address{$^1$Instituto de F\'{\i}sica de L\'{\i}quidos y Sistemas Biol\'ogicos,
 Grupo F\'{\i}sica del S\'olido, Casilla de Correo 565, La Plata 1900, Argentina}
\address{$^2$Departamento de F\'{\i}sica, CNEA, Avda. General Paz 1499
  y Constituyentes, 1650 San Martin, Argentina}
\address{$^3$Department of Physics, University of California, Davis CA 95616}

%\date{\today}
\maketitle

%\tightenlines

\begin{abstract}
Electron-hole mixing-induced fine structure
in alkaline earth hexaborides leads to 
lower energy (temperature) scales, and thus stronger tendency
toward an excitonic instability, than in
their doped counterparts ({\it viz.} Ca$_{1-x}$La$_{x}$B$_6$,
$x \approx 0.005)$, which 
are high Curie temperature, small moment ferromagnets.  
Comparison of Fermi surfaces and spectral distributions with de Haas -
van Alphen (dHvA), optical, transport, and tunneling data indicates that \sr
remains a fermionic semimetal down to (at least) 5 K, rather than 
forming an excitonic condensate.   
For the doped system the Curie temperature is higher than the 
degeneracy temperature.

\end{abstract}
\vskip 1cm
\footnoterule

%$^{\ddag}$ \parbox[t]{6in}{
%{\small E-mail: ~~pickett@physics.ucdavis.edu} } 
%\newpage

% for twocolumn activate the line below...
]

The observation of very low moment magnetism (0.07 $\mu_B$ per carrier)
in the very low carrier density doped semimetal 
Ca$_{0.995}$La$_{0.005}$B$_6$ (one carrier per $6\times 6\times6$
unit cells) at remarkably high temperature (T$_c$ =
600 K) by Young {\it et al.}\cite{young} has fueled
rapid investigation into possible mechanisms for this novel 
occurrence.\cite{zhit1,gorkov,balents}  The focus has been on some sort of
excitonic instability, possibly an excitonic condensate that breaks 
time reversal symmetry and thereby allows ferromagnetism.  What has not yet
drawn the same attention is the 
{\it undoped} system,
which is really the more suitable candidate for an
excitonic instability than the doped system.
The DB$_6$ hexaborides (D=Ca, Sr, Ba) would be semiconductors
with a band gap of 2 - 3 eV between bonding and antibonding B$_6^{2-}$
$2p$ bands except for a D atom $d$ band that dives through the gap
from above, with a minimum at the X point.  The
result is a very small band
overlap $\sim$90 meV at the three Brillouin zone (BZ) edge points X.

The study of instabilities in such semimetals goes back to Keldysh and
Kopaev\cite{KK} and des Cloizeaux.\cite{desCl}  In the presence of a 
residual screened
Coulomb interaction between the carriers and {\it sufficient nesting}, there
are possibilities of charge density wave (CDW) or spin density wave (SDW)
(`singlet' or `triplet') instabilities depending on the degree of nesting.
Zhitomirsky {\it et al.}\cite{zhit1} have suggested that magnetism can appear
as a result of consecutive CDW and SDW instabilities in a doped 
excitonic insulator, with polarization becoming allowed 
because both inversion and time-reversal symmetries are broken.
Barzykin and Gor'kov\cite{gorkov} contend that this model of
nested semimetals is instead
unstable to the appearance of a superstructure.
Balents and
Varma\cite{balents} have considered the relative stability of the various 
order parameters involving charge, spin, and X-point pockets and the
dependence on doping concentration $x$, and conclude that intrapocket
condensation is favored over interpocket pairing. 
 
Each of these discussions presumes the instability of the undoped system,
a question that has been revisited by Zhitomirsky and Rice.\cite{zhit2}
They emphasize the importance of interpocket scattering processes and 
reemphasize that the undoped hexaborides should be good candidates
for a condensed excitonic state.  What we show in this paper, by
comparison with band structure derived quantities with the considerable
data available\cite{ott} for \sr, is evidence
that \sr remains a fermionic semimetal as
described by conventional band theory down to at least 5 K, and perhaps
down to 0.5 K where a transition to a {\it more conductive} (not
insulating) phase has
been observed.  The electronic structure of the $x$=0.005 system is
in fact much simpler (in the absence of an excitonic instability) and
the doping level and observed Fermi surface volume\cite{dHvA} 
is consistent 
with a fully polarized ferromagnet (FM).

The excitonic instability is entirely dependent on the band structure. 
Some of the band characterist\-ics have been presented by 
Hasegawa and Yanase\cite{Hase}
and by Massidda {\it et al.}\cite{massidda} and crystal stability has been
studied by Ripplinger {\it et al.}\cite{ripp}
Here we address the band structure in more detail than heretofore, 
and find that the 
calculated energy scales are in excellent agreement with the variety of
data\cite{ott} on SrB$_6$.  
The fine structure that we
discuss applies to the {\it undoped} systems and affects the tendency
toward instability, and it can be tested with dHvA
data if the system remains fermionic at low temperature (T).  
We have calculated the band structures of these compounds
using the accurate, full potential
augmented plane wave method.\cite{wien}  In these hexaborides
the differences between the
local density approximation\cite{lda} and the generalized gradient
approximation\cite{gga} (GGA) are small even on the scale of the fine structure
we will be discussing, and we use the GGA results.  We discuss primarily
SrB$_6$ because of the more extensive experimental data,\cite{ott} 
and because when doped, SrB$_6$ also
shows (as does \ba) a FM phase similar to doped CaB$_6$, but at even higher 
temperature.\cite{fisk}  We use
the measured lattice constant of 4.20 \AA~and internal parameter
(B position along the cubic axis) of 0.203, the latter 
confirmed by our total energy
minimization.  We have checked that spin-orbit coupling has no noticable
effect on the results we discuss.

The two bands that overlap at the X=(100)$\pi/a$ point, shown in Fig. 1,
are characterized 
at the most basic level by 
longitudinal ($m_l$) and transverse ($m_t$) masses, given by
$m_l$={0.50, -2.13} and $m_t$={0.21, -0.20} relative to the free electron
mass (hole masses are negative).  Thermal masses ($m_t^2 m_l)^{1/3}$
are 0.28 and -0.44 respectively.  The band overlap (`negative gap') is 
90 meV in \sr.  Crossing of the electron ($e$) and hole ($h$) bands
only occurs precisely along the $\Delta$
= ($\xi$,0,0)$\frac{\pi}{a}$ and Z = 
(1,$\xi$,0)$\frac{\pi}{a}$ lines; elsewhere
coupling results in anticrossing.  This mixing results in the
two overlapping and intersecting ellipsoids (in the absence of coupling)
becoming two distinct surfaces as shown in Fig. 2: a circular lens
centered on $\Delta$ containing holes, 
and a ring (``napkin ring'') centered at X
containing electrons.  Each surface is made of pieces of each of the
two ellipsoids, and each contains an $e$ and a $h$ part.

\begin{figure}
\epsfclipon
\epsfxsize=5.0cm
\centerline{\epsffile{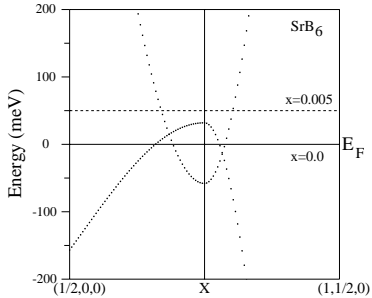}}
\caption{Calculated band structure near the X point.  The band crossings
occur only along these two lines, elsewhere there are anticrossings
(not shown).  The Fermi levels are shown as horizontal lines for
$x$=0 (solid) and $x$=0.005 (dashed).
The units of the axes are $\pi/a$.
\label{fig1}}
\end{figure}

We have used the condition that the number of electrons must equal the
number of holes to determine the intrinsic Fermi energy 
accurately.  The procedure
becomes quite delicate, requiring energy grids of spacing $\frac{1}{150}$
$\frac{\pi}{a}$ to represent the hybridization sufficiently accurately.
The result is that, whereas the ellipsoids at equal volumes would contain
$n_o$=2$\times 10^{-3}$ carriers, the six lenses and three rings 
contain only $n_e/2 = n_h/2$ =
2.7$\times 10^{-4}$ of the BZ volume.

The effect of band mixing is most easily seen in the drastic
effect it has on the density of states (DOS).  As pictured in Fig. 3,
instead of simple overlapping $(E - E_o)^{1/2}$
edges, there are rather
strong peaks near the points of mixing (due to flattened anticrossing
bands) with a deep minimum almost at the $x$=0 Fermi level.  This structure
makes the DOS at the Fermi level, N(E$_F$), depend much more strongly
on carrier concentration than might have been guessed.

\begin{figure}
\epsfclipon
\epsfxsize=5.5cm
\centerline{\epsffile{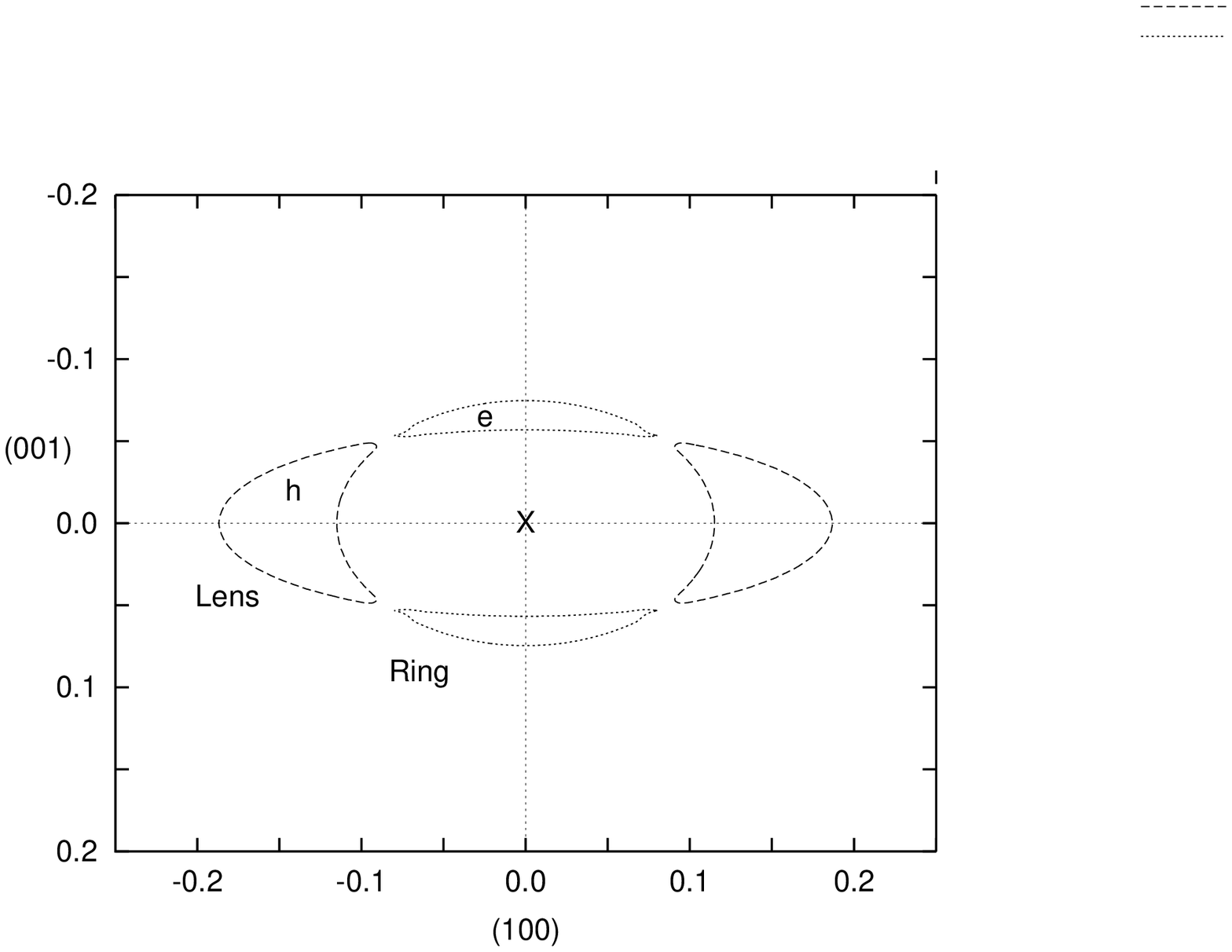}}
\caption{Cross section of the two Fermi surfaces of undoped SrB$_6$
centered on the X point.
The units of the axes are $\pi/a$.
The surfaces enclose electrons ($e$) and holes ($h$) surfaces as noted.
\label{fig2}}
\end{figure}

In free space this plasma ($n_e = n_h$) would be 
characterized by an electron gas 
parameter $r_s$ = 60, which would be an extremely low density plasma.
In a solid the effective mass (on average $|m^*| \sim 0.35$, see
above) and the background dielectric constant $\epsilon$ rescale the 
kinetic and potential energies, respectively, and alter the
effective density.  The value of $\epsilon$ has not been
reported.  Judging from the fact that SrB$_6$ would have an average direct
gap of around 3 eV except for the single conduction band that dips down
at X, and that this gap is similar to that of Si ($\epsilon$=12)
but the crystal is
less covalent and partially ionic, we estimate $\epsilon \sim 8$ for
the background dielectric constant.  This value
leads to plasma parameters
\begin{eqnarray}
a^*_B & = & a_B \epsilon \frac{m}{m^*} \approx 24 a_B 
                 \approx 13 \AA, \\ \nonumber
r^*_s & = & r_s \frac{a_B}{a^*_B} \approx 2.5 \\ \nonumber
E^*_R & = & E_R \frac{m^*}{m} \frac{1}{\epsilon^2} \approx 25~meV,
\end{eqnarray}
here $a_B$ is the Bohr radius.  The Rydberg E$_R^*$,
is the exciton binding energy that sets the scale of bandgaps
for which a low temperature excitonic instability is possible.
Using these values, the $e$ and $h$ densities of \sr
falls within the range of effective density of alkali metals, but
with an excitonic energy scale of E$_R^* \sim$ 25 meV that is two
orders of magnitude smaller than in alkali metals.  
The carrier plasma energy is changed little by this renormalization
]$(\epsilon^{-1} m/m^*)^{1/2} \sim 1/2$] and remains on the scale
of 100 meV.
The corresponding Fermi energies that determine the degeneracy temperature
are 20 meV 
for the holes and 15 meV for the electrons, making the degeneracy 
temperature T$_F\approx$250 K.  

\begin{figure}
\epsfclipon
\epsfysize=5cm
\centerline{\epsffile{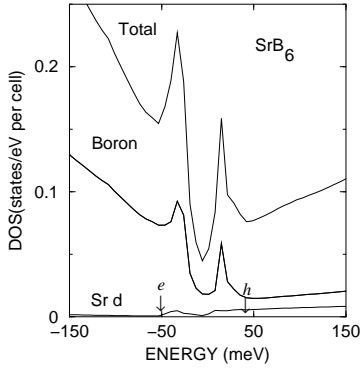}}
\caption{Density of states of SrB$_6$ near the Fermi level including
decomposition into B and Sr contributions.  The
band anticrossings lead to pronounced structure rather than a simple
overlapping to two square-root-like band edges.
The $e$ and $h$ band edges are marked.
\label{fig3}}
\end{figure}

Barring an excitonic instability,
at low temperature T$<<$T$_F$
dHvA oscillations will measure the
cross sectional areas and cyclotron masses of the extremal 
Fermi surface orbits perpendicular to 
the applied field.  For the field along a cubic axis, these Fermi surfaces
lead to five extremal orbits whose band masses we have calculated
from the energy derivative of the cross sectional area
$m^* = \frac{\hbar^2}{2\pi}dA/dE$.  The lens and ring cross sections
$O_{lens,0}$ and $O_{ring,0}$ lie in the plane of Fig. 2.  There are three
extremal orbits perpendicular to these (circling the $\Delta$ line):
$O_{lens,1}$, $O_{ring,in}$, and $O_{ring,out}$, where $in$ and $out$ 
denote the inner and outer ring orbits.

The areas and cyclotron band masses are given in Table I.  The values
of the three observed frequencies\cite{dHvA} are 30,
48, and 308 Tesla.  The first two are in the range of our values, and
in fact agree well with the two lens orbits.
These two orbits, both areas and masses, arise from the
mixing that eliminates the intersections of ellipsoids and creates new orbits
that can be seen in Fig. 2.  The 308 Tesla orbit 
is rather large to be accounted
for in terms of band structure results; even with magnetic breakdown,
the largest orbit (for the cigar ellipsoid) is only 196 Tesla.
It is interesting that the available data is consistent with ``missing'' 
rings, but it is also possible that ring orbits are simply more
difficult to observe.

The distinctive characters of the carriers may play an important role
in the behavior of the hexaborides,
{\it e.g.} by affecting matrix elements or, due to La doping on the 
alkaline earth sublattice, reducing the mean free path for the electrons
much more than for the holes.
The band that dips down through the gap has Sr $4d$ character (Fig. 3), 
and this
band forms the electron ellipsoids.  Thus the $e$ carriers are a
combination of
Sr $d$ and B $p$, while the $h$ carriers are purely bonding B
$p$ character.  Due to the mixing that rehybridizes the bands and
leads to the lenses and rings, each Fermi surface is roughly
half $e$-like and half $h$-like, and thus partly Sr $d$ as well as B $p$.

We now compare our results with available data on \sr.\cite{ott}  
The band structures
indicate a degeneracy scale E$^*$ 
of 15-20 meV (degeneracy temperature T$^* \sim$ 250 K), reduced from 600 K
by anticrossing electron and hole bands.  Interband transitions in the
infrared should peak at the separation of DOS peaks in Fig. 3,
near $\omega^* \sim$ 40 meV/$\hbar \sim$
320 cm$^{-1}$.  The experimental data indicate: 
\noindent (1) a change from metallic slope of resistivity 
$\rho$ to non-metallic 
slope at 250 K $\sim$ T$^*$, consistent with some change as the
system evolves from non-degenerate to degenerate (the transport
is not understood), 
\noindent (2) a rise in the optical conductivity (at low T) below 300 cm$^{-1}
\sim \omega^*$, but the peaking at 40-80 cm$^{-1}$ ({\it i.e.} below 10 meV)
may reflect k-conserving processes not identifiable in the DOS of Fig. 3,
\noindent (3) a broadening and weakening of this absorption peak
on the scale of 200-300 K $\sim$ T$^*$,
\noindent (4) a strong
onset of interband transitions at $\sim$2 eV, roughly consistent with
the band structure, 
\noindent (5) negligible electronic contribution to the specific heat
above 5 K; the band structure value would not be detectable,
\noindent (6) tunneling spectra show a conductance minimum at a 
negative bias of -40 meV $\approx \hbar \omega^*$, 
and strong T dependence of the spectrum at
-40 meV and at zero bias.\cite{tunnel}
All of these observations are consistent, almost quantitatively,
with the band overlap and mixing
that we calculate.

There are additional observations\cite{ott}: 
$\rho(T)$ shows structure at 50-60 K,
and drops very rapidly by 30\% below 0.5 K, this latter
temperature corresponding to
an energy scale of 40 $\mu$eV for which there is no immediate explanation.
There is excess specific heat
(above the lattice contribution) below 5 K, but with only a small 
integrated entropy of less than 0.05\% of $R~ln2$.
Except for this low temperature regime, observations seem consistent with a
fermionic plasma with band structure as described above.

Now we turn to the doped magnetic system. 
Doping of $x$=0.005 increases the
Fermi level by 50 meV as shown in Fig. 1, for which there is a 
single ellipsoidal Fermi surface
(at each of the X points) with eccentricity $\eta$=
 k$_{\parallel}$/k$_{\perp}$
= 1.6.  The hole surface vanishes near $x$=0.004.
For doped ($x$=0.005) CaB$_6$, the reported dHvA frequencies of 
350 T and 495 T \cite{dHvA} are consistent with such ellipsoids with
eccentricity $\eta$=495/350=1.4.  The required values of k$_{\perp}$
and k$_{\parallel}$ correspond to a doping level of 0.0090 for a
paramagnetic system; however, if the ellipsoids represent a fully polarized
electron gas, the corresponding doping level is 0.0045, very close to
the nominal doping level.  The reported
ordered moment of 0.07 $\mu_B$ per carrier might reflect large
magnetic fluctuations, noncollinearity of spins, orbital currents,
or phase separation in the magnetic phase.

The {\it origin} of the magnetism has had no explanation
except as an outgrowth of an excitonic undoped phase, which phase 
in SrB$_6$ seems not
to be supported by the data.
At $x$=0.005 and higher, the electronic structure of the paramagnetic
systems seems to be
very simple.
Inter-pocket $e-e$ nesting between inequivalent X 
points is the only $\omega
\rightarrow 0$ nesting (Q$\neq 0)$,
and is maximum between points on their equators.  
If the Fermi surfaces were spheres, $\omega$
= 0 nesting would
be perfect for Q=(1,1,0)$\pi/a$ wavevectors, tending to drive CDW or SDW
instabilities.  Due to the eccentricity ({\it i.e.} $m_t \neq m_l$), 
these scattering
processes are smeared by $\delta$Q$\sim 2\times 10^{-2} \pi/a$ for
$\omega$=0 or by $\delta \omega \sim 30$ meV at Q itself. 
Moreover,
the system is already magnetic above the degeneracy temperature, where
nesting is an irrelevant process.  At high T a 
nonzero -- but rather
small -- density of holes will be thermally excited, leading perhaps to a 
charge-unbalanced, quasiclassical $e-h$ plasma, but the high temperature
should also preclude formation of an exciton-fermion plasma.

To summarize, we have looked closely at the electronic structure of
undoped and doped divalent hexaborides, and compared them with the
available transport, thermal, optical, dHvA, and
tunneling data.  The behavior
of SrB$_6$ above 5 K is consistent with expectations, and
energy scales, obtained from the band structure.  The observed Fermi
surface of $x$=0.005 doped CaB$_6$ is most easily accounted for 
as the majority Fermi surface of a fully polarized electron gas,
but this interpretation is not readily reconciled with the measured tiny
moment.  

We acknowledge helpful communications with Z. Fisk, R. G. Goodrich, and J. W.
Allen during the course of this work, and comments on the manuscript by
R. Monnier.
This work was supported by an NSF-CONICET International Collaboration.
W. E. P. was supported by National Science Foundation (NSF) Grant No.
DMR-9802076.
Ruben Weht acknowledges support from
Fundaci\'on Antorchas Grants No. A-13622/1-103 and A-13661/1-27.
Much of this work was done when the authors were at the Institute of
Theoretical Physics in Santa Barbara, which is supported by 
NSF Grant No. PHY 94-07194. 

% References

\begin{table}
\caption{Calculated Fermi surface properties of \sr.  F denotes the dHvA
frequency and $m^*$ denotes the band
masses. See text for notation.}
\begin{tabular}{lccccc}
 & O$_{lens,0}$ & O$_{lens,1}$ & O$_{ring,0}$ & O$_{ring,out}$ & O$_{ring,in}$\\\hline
F(Tesla) &  33  &     46       &     10       &     98     &   61      \\
$m^*_b/m_o$ & -0.35 & $\sim 0.2$ &    0.35      &   0.2    &   -0.2  \\
\hline
\end{tabular}
\label{table1}
\end{table}
\end{document}